\UseRawInputEncoding
\documentclass[journal]{IEEEtran}
\ifCLASSINFOpdf
\else
   \usepackage{graphicx}
\fi
\usepackage{url}

\usepackage{graphicx}

\begin{document}

\title{Target Clustering Based Multi-Bernoulli Filter for Superpositional Sensors}

\author{Wang Sen
\thanks{This paragraph of the first footnote will contain the date on which you submitted your paper for review. \emph{(Corresponding author: Wang Sen)}}
\thanks{The authors are with National University of Defense Technology, Xi'an 710106, China (e-mail: wangsen11@nudt.edu.cn)}}

\markboth{Journal of \LaTeX\ Class Files, Vol. 14, No. 8, August 2015}
{Shell \MakeLowercase{\textit{et al.}}: Bare Demo of IEEEtran.cls for IEEE Journals}
\maketitle

\begin{abstract}
The sensor whose output is a function of the sum of contributions from targets present in the surveillance area is called superpositional sensor. In this letter, target clustering based multi-Bernoulli filter for superpositional sensors is proposed. Targets are clustered according to the set of resolution cells illuminated by them. Single target posterior density is strictly derived, and densities of all the targets are combined to a approximate multi-target posterior, which makes the multi-Bernoulli density is conjugate with respect to the likelihood of superpositional sensors. The Gaussian implementation of the proposed algorithm is also presented, where the multidimensionality and the nonlinearity of update equation are handled by sigma point transformation. The simulation results illustrate that the proposed algorithm is effective confronted with the interaction of multiple targets and long term overlapping of two targets.
\end{abstract}

\begin{IEEEkeywords}
Multi-Bernoulli filter, superpositional sensors, target clustering, sigma point transformation, track before detect (TBD).
\end{IEEEkeywords}

\IEEEpeerreviewmaketitle

\section{Introduction}
\IEEEPARstart{M}{ulti-target} tracking (MTT) is concerned with estimating the number of targets and their individual states from a sequence of measurements. The detect before track (DBT) method approaches point measurements obtained by a detection process, and may not be adequate for applications with low signal to noise ratio (SNR). The track before detect (TBD) paradigm, which directly processes raw measurements, can make use of all information contained in image observations to improve tracking performance. When tracking multiple closely spaced targets, regions of target influence in image observations are overlapping, and the sensor output is a function of the sum of contributions from individual targets, which is called the superpositional sensor. Examples belonging to superpositional sensors are direction of arrival estimation for antenna array \cite{Dong2023}, acoustic amplitude sensor \cite{Saucan2017}, Doppler radar \cite{Wenhao2022}, and radio frequency tomography \cite{Ri2023}.

For superpositional sensors, the update equation of cardinalized probability hypothesis density (CPHD) filter is derived \cite{Mahler2009}, and it is theoretically rigorous but computationally intractable. Based on probability hypothesis density (PHD) approximation, the additive likelihood moment (ALM) filter and its particle implementation are designed \cite{Thouin2011}. And then, the ALM filter is generalized to CPHD filtering recursion \cite{Ronald2012,Nannuru2013}. Unlike the PHD/CPHD recursions, the multi-Bernoulli recursion approximately propagates the full multi-target distribution, and avoids postprocessing procedures such as clustering to obtain individual target state. Under the assumption that the regions of the observation influenced by individual targets do not overlap, the multi-Bernoulli-TBD (MB-TBD) filters for various prior distributions are derived \cite{Vo2010}. In the context of superpositional measurements, the multi-Bernoulli prior is not conjugate, and approximation is inevitable. In \cite{Nannuru2013a,Nannuru2013b}, the multi-Bernoulli posterior distribution is propagated by updating predefined conditional PHD. In \cite{Kim2021}, to obtain the single target likelihood, other interacting targets are treated as interferences. In \cite{Davies2022}, the predicted measurement information of each target is shared with the rest of the targets.

In this letter, target clustering based multi-Bernoulli (TC-MB) filter for superpositional sensors is proposed. The set of resolution cells illuminated by the target is defined. Accordingly, the targets are clustered, and each cluster will be updated jointly. The single target posterior density is found by integrating out the other targets, and it is a Bernoulli density. Combining all the Bernoulli densities produces a multi-Bernoulli density, which can approximate the real multi-target posterior. In the Gaussian implementation, the multidimensionality and the nonlinearity of update equation are handled by sigma point transformation.

\section{Background}
The multi-target motion model based on random finite sets (RFS) comprehensively considers target birth, death, spawning, and motion. Suppose the multi‐target state at time $k-1$ is represented as a RFS ${{\bf{X}}_{k - 1}} = \left\{ {{\bf{x}}_{k - 1}^1,{\bf{x}}_{k - 1}^2, \cdots ,{\bf{x}}_{k - 1}^{{M_{k - 1}}}} \right\}$, where ${M_{k - 1}}$ is the number of targets, regardless of spawned targets, the multi-target state at time $k$ is ${{\bf{X}}_k} =  \cup S\left( {{\bf{x}}_{k - 1}^1} \right) \cdots  \cup S\left( {{\bf{x}}_{k - 1}^{{M_{k - 1}}}} \right) \cup {\Gamma _k}$, where $S\left( {{\bf{x}}_{k - 1}^m} \right)$ is the state of target ${\bf{x}}_{k - 1}^m$ at time $k$, $m = 1,2, \cdots ,{M_{k - 1}}$, determined by the survival probability ${p_S}\left( {{\bf{x}}_{k - 1}^m} \right)$ and single target Markov transition density $f\left( {{\bf{x}}_k^m|{\bf{x}}_{k - 1}^m} \right)$, and ${\Gamma _k}$ is the RFS of completely new targets.

At time $k$, the measurement provided by a superpositional sensor is a random array ${{\bf{Z}}_k} = \left[ {z_k^n} \right]$, where $n = 1,2, \cdots ,N$, with $N$ being the number of cells. One resolution cell could be illuminated by multiple targets. The set of targets contributing to cell $n$ is denoted as $\Theta \left( n \right)$. Consequently, the measurement $z_k^n$ has the following expression
\begin{equation}
\label{measurementequation}
z_k^n = \left\{ {\begin{array}{*{20}{l}}
{\sum\limits_{{\bf{x}} \in \Theta \left( n \right)} {{h^n}\left( {\bf{x}} \right)}  + {w^n}}&{\Theta \left( n \right) \ne \emptyset }\\
{{w^n}}&{\Theta \left( n \right) = \emptyset }
\end{array}} \right.
\end{equation}
where ${h^n}\left( {\bf{x}} \right)$ represents the contribution of target ${\bf{x}}$ to the measurement of cell $n$, and ${w^n}$ is zero-mean Gaussian white noise with covariance $R$.

Define ${{\bf{Z}}_{1:k}} = \left[ {{{\bf{Z}}_1},{{\bf{Z}}_2}, \cdots ,{{\bf{Z}}_k}} \right]$, the posterior probability density $p\left( {{{\bf{X}}_k}|{{\bf{Z}}_{1:k}}} \right)$ can be computed using Bayes rule
\begin{equation}
p\left( {{{\bf{X}}_k}|{{\bf{Z}}_{1:k - 1}}} \right) = \int {p\left( {{{\bf{X}}_k}|{{\bf{X}}_{k - 1}}} \right)p\left( {{{\bf{X}}_{k - 1}}|{{\bf{Z}}_{1:k - 1}}} \right)\delta {{\bf{X}}_{k - 1}}}
\end{equation}
\begin{equation}
\label{update}
p\left( {{{\bf{X}}_k}|{{\bf{Z}}_{1:k}}} \right) = \frac{{p\left( {{{\bf{Z}}_k}|{{\bf{X}}_k}} \right)p\left( {{{\bf{X}}_k}|{{\bf{Z}}_{1:k - 1}}} \right)}}{{\int {p\left( {{{\bf{Z}}_k}|{{\bf{X}}_k}} \right)p\left( {{{\bf{X}}_k}|{{\bf{Z}}_{1:k - 1}}} \right)\delta {{\bf{X}}_k}} }}
\end{equation}

A multi-Bernoulli RFS ${\bf{X}}$ is the disjoint union of a fixed number of independent Bernoulli RFSs, and has density
\begin{equation}
{p_{{\rm{MB}}}}\left( {\bf{X}} \right) = \sum\limits_{{{\bf{X}}^1} \uplus  \cdots  \uplus {{\bf{X}}^M} = {\bf{X}}} {\prod\limits_{m = 1}^M {{f^m}\left( {{{\bf{X}}^m}} \right)} }
\end{equation}
where $\uplus$ denotes disjoint union, and the Bernoulli RFS ${{\bf{X}}^m}$ has density ${f^m}$ with existence probability ${r^m}$ and existence-conditioned probability density function (PDF) ${p^m}\left( {\bf{x}} \right)$, $m = 1, \cdots ,M$.

\section{Target Clustering Based Multi-Bernoulli Filter}
The proposed target clustering based multi-Bernoulli filter for superpositional sensors propagates over time the parameters of the multi-Bernoulli RFS representing the multi-target state.
\subsection{Multi-Bernoulli Filter Prediction}
The superpositional sensors do not change the multi-Bernoulli prediction equations \cite{Mahler2007,Vo2010}. Suppose the posterior multi-Bernoulli density at time $k-1$ is characterized by parameters $\left( {r_{k - 1|k - 1}^m,p_{k - 1|k - 1}^m\left( {\bf{x}} \right)} \right),m = 1, \cdots ,{M_{k - 1|k - 1}}$ and completely new targets at time $k$ is characterized by parameters $\left( {r_k^{b,m},p_k^{b,m}\left( {\bf{x}} \right)} \right),m = 1, \cdots ,M_k^b$, the prior multi-Bernoulli density at time $k$ can be written as
\begin{equation}
\label{prior}
\begin{array}{l}
p\left( {{{\bf{X}}_k}|{{\bf{Z}}_{1:k - 1}}} \right) = \\
\sum\limits_{{\bf{X}}_{k|k - 1}^1 \uplus  \cdots  \uplus {\bf{X}}_{k|k - 1}^{{M_{k|k - 1}}} = {{\bf{X}}_k}} {\prod\limits_{m = 1}^{{M_{k|k - 1}}} {f_{k|k - 1}^m\left( {{\bf{X}}_{k|k - 1}^m} \right)} } 
\end{array}
\end{equation}
where ${M_{k|k - 1}} = {M_{k - 1|k - 1}} + M_k^b$ is the number of Bernoulli components, and the Bernoulli RFS ${\bf{X}}_{k|k - 1}^m$ has density
\begin{equation}
f_{k|k - 1}^m\left( {{\bf{X}}_{k|k - 1}^m} \right) = \left\{ {\begin{array}{*{20}{l}}
{1 - r_{k|k - 1}^m}&{{\bf{X}}_{k|k - 1}^m = \emptyset }\\
{r_{k|k - 1}^mp_{k|k - 1}^m\left( {\bf{x}} \right)}&{{\bf{X}}_{k|k - 1}^m = \left\{ {\bf{x}} \right\}}\\
0&{\left| {{\bf{X}}_{k|k - 1}^m} \right| > 1}
\end{array}} \right.
\end{equation}
For the surviving Bernoulli components, $m = 1, \cdots ,{M_{k - 1|k - 1}}$, the parameters are
\begin{equation}
r_{k|k - 1}^m = r_{k - 1|k - 1}^m\left\langle {p_{k - 1|k - 1}^m,{p_S}} \right\rangle
\end{equation}
\begin{equation}
p_{k|k - 1}^m\left( {\bf{x}} \right) = \frac{{\int {p_{k - 1|k - 1}^m\left( {\bf{x}} \right){p_S}\left( {\bf{x}} \right)f\left( { \cdot |{\bf{x}}} \right)d{\bf{x}}} }}{{\left\langle {p_{k - 1|k - 1}^m,{p_S}} \right\rangle }}
\end{equation}
where $\left\langle {a,b} \right\rangle  = \int {a\left( {\bf{x}} \right)b\left( {\bf{x}} \right)d{\bf{x}}}$ represents the inner product of $a\left( {\bf{x}} \right)$ and $b\left( {\bf{x}} \right)$. For Bernoulli components of completely new targets, $m = {M_{k - 1|k - 1}} + 1, \cdots ,{M_{k|k - 1}}$, the parameters are
\begin{equation}
\label{existenceprobabilitynew}
r_{k|k - 1}^m = r_k^{b,m - {M_{k - 1|k - 1}}}
\end{equation}
\begin{equation}
p_{k|k - 1}^m\left( {\bf{x}} \right) = p_k^{b,m - {M_{k - 1|k - 1}}}\left( {\bf{x}} \right)
\end{equation}

\subsection{Target Clustering}
\label{TargetClustering}
For the superpositional measurement model, the targets are interacting with each other in the measurement space inevitably, and the corresponding single target likelihood is not available. In the proposed algorithm, the targets are clustered, and each cluster will be updated jointly.

The set of resolution cells illuminated by target ${\bf{X}}_{k|k - 1}^m$ is denoted as $\Phi \left( {{\bf{X}}_{k|k - 1}^m} \right)$, for example $\Phi \left( {{\bf{X}}_{k|k - 1}^m} \right)$ could be the set of cells whose centers fall within a certain distance from the position of the target. If there exists resolution cell ${n^ * }$ being illuminated by target ${\bf{X}}_{k|k - 1}^{{m_1}}$ and ${\bf{X}}_{k|k - 1}^{{m_2}}$, namely ${n^ * } \in \Phi \left( {{\bf{X}}_{k|k - 1}^{{m_1}}} \right) \cap \Phi \left( {{\bf{X}}_{k|k - 1}^{{m_2}}} \right)$, target ${\bf{X}}_{k|k - 1}^{{m_1}}$ and ${\bf{X}}_{k|k - 1}^{{m_2}}$ are clustered into one cluster, and they are called the interacting targets; otherwise, target ${\bf{X}}_{k|k - 1}^{{m_1}}$ and ${\bf{X}}_{k|k - 1}^{{m_2}}$ are clustered into two different clusters, and they are called the well-separated targets. Accordingly, multiple targets ${{\bf{X}}_k}$ will be divided into ${C_k}$ clusters
\begin{equation}
\label{clusters}
{{\bf{X}}_k} = {\bf{X}}_k^{\left( 1 \right)} \uplus  \cdots {\bf{X}}_k^{\left( c \right)} \uplus  \cdots {\bf{X}}_k^{\left( {{C_k}} \right)}
\end{equation}
One cluster may contain more than two targets. For example, if $\Phi \left( {{\bf{X}}_{k|k - 1}^{{m_1}}} \right) \cap \Phi \left( {{\bf{X}}_{k|k - 1}^{{m_2}}} \right) \ne \emptyset$ and $\Phi \left( {{\bf{X}}_{k|k - 1}^{{m_1}}} \right) \cap \Phi \left( {{\bf{X}}_{k|k - 1}^{{m_3}}} \right) \ne \emptyset$, target ${\bf{X}}_{k|k - 1}^{{m_1}}$, ${\bf{X}}_{k|k - 1}^{{m_2}}$, and ${\bf{X}}_{k|k - 1}^{{m_3}}$ belong to one cluster. One cluster also may contain one target, when it is well separated from others. Naturally, the set of resolution cells illuminated by targets in ${\bf{X}}_k^{\left( c \right)}$ is defined as
\begin{equation}
\Phi \left( {{\bf{X}}_k^{\left( c \right)}} \right) \buildrel \Delta \over = \bigcup\limits_{{\bf{X}}_{k|k - 1}^m \in {\bf{X}}_k^{\left( c \right)}} {\Phi \left( {{\bf{X}}_{k|k - 1}^m} \right)} ,c = 1, \cdots ,{C_k}
\end{equation}
and the set of resolution cells illuminated by targets at time $k$ is defined as
\begin{equation}
\Phi \left( {{{\bf{X}}_k}} \right) \buildrel \Delta \over = \bigcup\limits_{{\bf{X}}_{k|k - 1}^m \in {{\bf{X}}_k}} {\Phi \left( {{\bf{X}}_{k|k - 1}^m} \right)}
\end{equation}
The multi-target likelihood function can be written as
\begin{equation}
\label{likelihood}
\begin{array}{l}
p\left( {{{\bf{Z}}_k}|{{\bf{X}}_k}} \right) = \prod\limits_{n = 1}^N {{\rm{{\cal N}}}\left( {z_k^n;0,R} \right)}  \times \\
\prod\limits_{c = 1}^{{C_k}} {\prod\limits_{n \in \Phi \left( {{\bf{X}}_k^{\left( c \right)}} \right)} {\frac{{{\rm{{\cal N}}}\left( {z_k^n;\sum\limits_{{\bf{x}} \in \Theta \left( n \right)} {{h^n}\left( {\bf{x}} \right)} ,R} \right)}}{{{\rm{{\cal N}}}\left( {z_k^n;0,R} \right)}}} } 
\end{array}
\end{equation}
where ${\rm{{\cal N}}}\left( { \cdot ;\mu ,\Sigma } \right)$ denotes the probability density function of the Gaussian distribution with mean $\mu$ and covariance matrix $\Sigma$.

\subsection{Multi-Bernoulli Filter Update}
Substituting (\ref{prior}) and (\ref{likelihood}) into (\ref{update}) produces the multi-target posterior density
\begin{equation}
\label{posterior}
p\left( {{{\bf{X}}_k}|{{\bf{Z}}_{1:k}}} \right) \propto \sum\limits_{{\bf{X}}_{k|k - 1}^1 \uplus  \cdots  \uplus {\bf{X}}_{k|k - 1}^{{M_{k|k - 1}}} = {{\bf{X}}_k}} {\prod\limits_{c = 1}^{{C_k}} {p\left( {{\bf{X}}_k^{\left( c \right)}|{{\bf{Z}}_{1:k}}} \right)} }
\end{equation}
where the posterior density of targets in ${\bf{X}}_k^{\left( c \right)}$ is
\begin{equation}
\label{clusterposterior}
\begin{array}{l}
p\left( {{\bf{X}}_k^{\left( c \right)}|{{\bf{Z}}_{1:k}}} \right) \propto \\
\prod\limits_{{\bf{X}}_{k|k - 1}^m \in {\bf{X}}_k^{\left( c \right)}} {f_{k|k - 1}^m\left( {{\bf{X}}_{k|k - 1}^m} \right)} \prod\limits_{n \in \Phi \left( {{\bf{X}}_k^{\left( c \right)}} \right)} {\frac{{{\rm{{\cal N}}}\left( {z_k^n;\sum\limits_{{\bf{x}} \in \Theta \left( n \right)} {{h^n}\left( {\bf{x}} \right)} ,R} \right)}}{{{\rm{{\cal N}}}\left( {z_k^n;0,R} \right)}}} 
\end{array}
\end{equation}
Equation (\ref{posterior}) indicates that the multi-target posterior density is the product of the posterior density of targets in each cluster.

The multi-target likelihood ratio in (\ref{clusterposterior}) can be expressed as
\begin{equation}
{L^{\left( c \right)}}\left( {{b_1}, \cdots ,{b_{\left| {{\bf{X}}_k^{\left( c \right)}} \right|}}} \right) = \prod\limits_{n \in \Phi \left( {{\bf{X}}_k^{\left( c \right)}} \right)} {\frac{{{\rm{{\cal N}}}\left( {z_k^n;\sum\limits_{{\bf{x}} \in \Theta \left( n \right)} {{h^n}\left( {\bf{x}} \right)} ,R} \right)}}{{{\rm{{\cal N}}}\left( {z_k^n;0,R} \right)}}}
\end{equation}
All the arguments in ${L^{\left( c \right)}}$ are booleans, representing the existence of targets. For example, ${b_1} = 1$ indicates that the first target in ${\bf{X}}_k^{\left( c \right)}$ exists, while ${b_1} = 0$ means that the first target in ${\bf{X}}_k^{\left( c \right)}$ does not exist.

The posterior density of target ${\bf{X}}_k^m$ (${\bf{X}}_{k|k - 1}^m$) is found by integrating out the other targets
\begin{equation}
\label{singletargetposteriordensity}
p\left( {{\bf{X}}_k^m|{{\bf{Z}}_{1:k}}} \right) = \int {p\left( {{\bf{X}}_k^{\left( c \right)}|{{\bf{Z}}_{1:k}}} \right)\delta {\bf{X}}_k^{\left( c \right)\backslash m}}
\end{equation}
where ${\bf{X}}_k^{\left( c \right)\backslash m} = {\bf{X}}_k^{\left( c \right)}\backslash {\bf{X}}_k^m$.

The single target posterior density (\ref{singletargetposteriordensity}) is a Bernoulli density, and the multi-target posterior density (\ref{posterior}) can be approximated by the following multi-Bernoulli density
\begin{equation}
p\left( {{{\bf{X}}_k}|{{\bf{Z}}_{1:k}}} \right) \approx \sum\limits_{{\bf{X}}_k^1 \uplus  \cdots  \uplus {\bf{X}}_k^{{M_k}} = {{\bf{X}}_k}} {\prod\limits_{m = 1}^{{M_k}} {p\left( {{\bf{X}}_k^m|{{\bf{Z}}_{1:k}}} \right)} }
\end{equation}
where ${M_k} = {M_{k|k - 1}}$. Thus, it is achieved that the multi-Bernoulli density is conjugate with respect to the likelihood of superpositional sensors.

\section{Gaussian Implementation}
This section describes the Gaussian implementation of the proposed algorithm.
\subsection{Prediction}
The Gaussian implementation of prediction is obtained when there are a constant survival probability ${p_S}\left( {{\bf{x}}_{k - 1}^m} \right) \equiv {p_S}$ and Gaussian/linear single target Markov transition density
\begin{equation}
f\left( {{\bf{x}}_k^m|{\bf{x}}_{k - 1}^m} \right) = {\rm{{\cal N}}}\left( {{\bf{x}}_k^m;{\bf{Fx}}_{k - 1}^m,{\bf{Q}}} \right)
\end{equation}
where ${\bf{F}}$ and ${\bf{Q}}$ are the state transition matrix and the process noise covariance, respectively. Suppose the posterior state density of the Bernoulli RFS ${\bf{X}}_{k - 1|k - 1}^m,m = 1, \cdots ,{M_{k - 1|k - 1}}$ is approximated as
\begin{equation}
p_{k - 1|k - 1}^m\left( {\bf{x}} \right) = {\rm{{\cal N}}}\left( {{\bf{x}};{\bf{\bar x}}_{k - 1|k - 1}^m,{\bf{P}}_{k - 1|k - 1}^m} \right)
\end{equation}
the prior mean and covariance of surviving targets at time $k$ are provided by the Kalman filter prediction
\begin{equation}
{\bf{\bar x}}_{k|k - 1}^m = {\bf{F\bar x}}_{k - 1|k - 1}^m
\end{equation}
\begin{equation}
{\bf{P}}_{k|k - 1}^m = {\bf{FP}}_{k - 1|k - 1}^m{{\bf{F}}^{\rm{T}}} + {\bf{Q}}
\end{equation}
which constitute the prior state density
\begin{equation}
p_{k|k - 1}^m\left( {\bf{x}} \right) = {\rm{{\cal N}}}\left( {{\bf{x}};{\bf{\bar x}}_{k|k - 1}^m,{\bf{P}}_{k|k - 1}^m} \right)
\end{equation}
and the prior existence probability is
\begin{equation}
r_{k|k - 1}^m = r_{k - 1|k - 1}^m{p_S}
\end{equation}

\subsection{Target Clustering}
In the implementation of target clustering, the set of resolution cells illuminated by target ${\bf{X}}_{k|k - 1}^m$ can be specified by the prior mean ${\bf{\bar x}}_{k|k - 1}^m$ when it exists, and be defined as
\begin{equation}
\Phi \left( {{\bf{X}}_{k|k - 1}^m} \right) = \Phi \left( {{\bf{\bar x}}_{k|k - 1}^m} \right)
\end{equation}
According to \ref{TargetClustering}, all the prior targets can be divided into ${C_k}$ clusters, seen (\ref{clusters}).

For clarity and conciseness, time index and cluster index of prior parameters are omitted. Specifically, the number of targets in ${\bf{X}}_k^{\left( c \right)}$ is expressed as $M = \left| {{\bf{X}}_k^{\left( c \right)}} \right|$, the Bernoulli RFS constituting ${\bf{X}}_k^{\left( c \right)}$ is expressed as ${{\bf{X}}^m},m = 1, \cdots ,M$, and the corresponding existence probability, state density, mean, and covariance are expressed as ${r^m}$, ${p^m}\left( {{{\bf{x}}^m}} \right)$, ${{{\bf{\bar x}}}^m}$, and ${{\bf{P}}^m}$, respectively. Therefore, (\ref{clusterposterior}) can be expressed as
\begin{equation}
\label{clusterposterior2}
\begin{array}{l}
p\left( {{\bf{X}}_k^{\left( c \right)}|{{\bf{Z}}_{1:k}}} \right) \propto \\
\left\{ {\begin{array}{*{20}{l}}
{\left( {1 - {r^1}} \right)\left( {1 - {r^2}} \right) \cdots \left( {1 - {r^M}} \right){L^{\left( c \right)}}\left( {0,0, \cdots ,0} \right)}\\
{\left( {1 - {r^1}} \right)\left( {1 - {r^2}} \right) \cdots {r^M}{p^M}\left( {{{\bf{x}}^M}} \right){L^{\left( c \right)}}\left( {0,0, \cdots ,1} \right)}\\
 \vdots \\
{{r^1}{p^1}\left( {{{\bf{x}}^1}} \right){r^2}{p^2}\left( {{{\bf{x}}^2}} \right) \cdots {r^M}{p^M}\left( {{{\bf{x}}^M}} \right){L^{\left( c \right)}}\left( {1,1, \cdots ,1} \right)}
\end{array}} \right.
\end{array}
\end{equation}
The condition of piecewise function (\ref{clusterposterior2}) is omitted, because it can be easily inferred by the arguments in ${L^{\left( c \right)}}$.

\subsection{Update}
The key of update is to obtain posterior target parameters using set integral (\ref{singletargetposteriordensity}), which is computationally intractable. Using (\ref{clusterposterior2}), (\ref{singletargetposteriordensity}) can be converted into vector integral
\begin{equation}
\label{singletargetposteriordensity2}
\begin{array}{l}
p\left( {{\bf{X}}_k^m|{{\bf{Z}}_{1:k}}} \right) \propto \\
\left\{ {\begin{array}{*{20}{l}}
{\left( {1 - {r^1}} \right)\left( {1 - {r^2}} \right) \cdots \left( {1 - {r^M}} \right){L^{\left( c \right)}}\left( {0,0, \cdots ,0} \right)}\\
{\int {\left( {1 - {r^1}} \right)\left( {1 - {r^2}} \right) \cdots {r^M}{p^M}\left( {{{\bf{x}}^M}} \right){L^{\left( c \right)}}\left( {0,0, \cdots ,1} \right)d{{\bf{x}}^M}} }\\
 \vdots \\
{\int {{r^1}{p^1}\left( {{{\bf{x}}^1}} \right) \cdots {r^M}{p^M}\left( {{{\bf{x}}^M}} \right){L^{\left( c \right)}}\left( {1,1, \cdots ,1} \right)d{{\bf{x}}^1} \cdots d{{\bf{x}}^M}} }
\end{array}} \right.
\end{array}
\end{equation}
It is worth noting that there is no $d{{\bf{x}}^m}$ in (\ref{singletargetposteriordensity2}). Vector integral
(\ref{singletargetposteriordensity2}) still encounters difficulties of the multidimensionality and the nonlinearity, and this subsection uses sigma point transformation to implement it.

The implementation is demonstrated with the last piece of (\ref{singletargetposteriordensity2}), and the rest pieces have simpler computations. Several sigma points and weights are generated from the prior state density, and can approximate it as following
\begin{equation}
\label{sigmapointsapproximation}
\begin{array}{l}
{p^u}\left( {{{\bf{x}}^u}} \right) = {\rm{{\cal N}}}\left( {{{\bf{x}}^u};{{{\bf{\bar x}}}^u},{{\bf{P}}^u}} \right) \approx \sum\limits_{{i_u} = 1}^{{N_S}} {{w_{{i_u}}}\delta \left( {{{\bf{x}}^u} - {\bf{\chi }}_{{i_u}}^u} \right)} \\
u = 1, \cdots ,m - 1,m + 1, \cdots ,M
\end{array}
\end{equation}
where ${N_S}$ is the number of sigma points. The multi-target likelihood ratio in the last piece of (\ref{singletargetposteriordensity2}) can be represented as a function of prior state variables
\begin{equation}
\label{multitargetlikelihoodratio}
l\left( {{{\bf{x}}^1}, \cdots ,{{\bf{x}}^m}, \cdots ,{{\bf{x}}^M}} \right) \buildrel \Delta \over = {L^{\left( c \right)}}\left( {1,1, \cdots ,1} \right)
\end{equation}
Substituting (\ref{sigmapointsapproximation}) and (\ref{multitargetlikelihoodratio}) into the last piece of (\ref{singletargetposteriordensity2}) produces
\begin{equation}
\sum\limits_{{i_1} = 1}^{{N_S}} { \cdots \sum\limits_{{i_M} = 1}^{{N_S}} {{r^1}{w_{{i_1}}} \cdots {r^M}{w_{{i_M}}}l\left( {{\bf{\chi }}_{{i_1}}^1, \cdots ,{\bf{\chi }}_{{i_M}}^M} \right)\delta \left( {{{\bf{x}}^m} - {\bf{\chi }}_{{i_m}}^m} \right)} }
\end{equation}
which is a function of ${{\bf{x}}^m}$. Similarly, half pieces of (\ref{singletargetposteriordensity2}) yield a function of ${{\bf{x}}^m}$, and the other half pieces yield a scalar. Suppose the sum of the former is denoted as $\sum\nolimits_{i = 1}^{{N_S}} {{{w'}_i}\delta \left( {{{\bf{x}}^m} - {\bf{\chi }}_i^m} \right)}$, and the sum of the latter is denoted as ${r'}$, then the posterior density (\ref{singletargetposteriordensity2}) has the form of Bernoulli density
\begin{equation}
p\left( {{\bf{X}}_k^m|{{\bf{Z}}_{1:k}}} \right) = \left\{ {\begin{array}{*{20}{l}}
{1 - \frac{{\sum\nolimits_{j = 1}^{{N_S}} {{{w'}_j}} }}{{r' + \sum\nolimits_{j = 1}^{{N_S}} {{{w'}_j}} }}}&{{\bf{X}}_k^m = \emptyset }\\
{\sum\limits_{i = 1}^{{N_S}} {\frac{{{{w'}_i}}}{{r' + \sum\nolimits_{j = 1}^{{N_S}} {{{w'}_j}} }}\delta \left( {{{\bf{x}}^m} - {\bf{\chi }}_i^m} \right)} }&{{\bf{X}}_k^m = \left\{ {{{\bf{x}}^m}} \right\}}
\end{array}} \right.
\end{equation}
Therefore, posterior parameters including existence probability, mean, and covariance of the Bernoulli RFS ${\bf{X}}_k^m$ are
\begin{equation}
r_{k|k}^m = \frac{{\sum\nolimits_{j = 1}^{{N_S}} {{{w'}_j}} }}{{r' + \sum\nolimits_{j = 1}^{{N_S}} {{{w'}_j}} }}
\end{equation}
\begin{equation}
{\bf{\bar x}}_{k|k}^m = \frac{{\sum\nolimits_{i = 1}^{{N_S}} {{{w'}_i}{\bf{\chi }}_i^m} }}{{\sum\nolimits_{i = 1}^{{N_S}} {{{w'}_i}} }}
\end{equation}
\begin{equation}
{\bf{P}}_{k|k}^m = \frac{{\sum\nolimits_{i = 1}^{{N_S}} {{{w'}_i}\left( {{\bf{\chi }}_i^m - {\bf{\bar x}}_{k|k}^m} \right){{\left( {{\bf{\chi }}_i^m - {\bf{\bar x}}_{k|k}^m} \right)}^{\mathop{\rm T}\nolimits} }} }}{{\sum\nolimits_{i = 1}^{{N_S}} {{{w'}_i}} }}
\end{equation}

\section{Simulation}
In this section, the performance of the proposed target clustering based multi-Bernoulli (TC-MB) filter for superpositional sensors is investigated and compared with the multi-Bernoulli-TBD (MB-TBD) filter \cite{Vo2010} in multi-target tracking application.

In the simulations, the sensor field of view is a two-dimensional region $\left[ {0,128} \right] \times \left[ {0,128} \right]$, and the total time of simulation is $K = 70$. There are five targets present in the sensor field of view, and single target state is defined as ${\bf{x}} = \left[ {x;\dot x;y;\dot y} \right]$, representing target position and velocity along the x-axis and y-axis. The target motion is modeled by a nearly constant velocity model
\begin{equation}
{{\bf{x}}_k} = {\bf{F}}{{\bf{x}}_{k - 1}} + {{\bf{n}}_{k - 1}}
\end{equation}
\begin{equation}
{\bf{F}} = {{\bf{I}}_2} \otimes \left[ {\begin{array}{*{20}{c}}
1&T\\
0&1
\end{array}} \right]
\end{equation}
where ${{\bf{I}}_2}$ is the $2 \times 2$ identity matrix, $T = 1$ is the sampling period, $ \otimes $ represents the Kronecker product, and the covariance of process noise ${{\bf{n}}_{k - 1}}$ is
\begin{equation}
{\bf{Q}} = {10^{ - 4}} \times {{\bf{I}}_2} \otimes \left[ {\begin{array}{*{20}{c}}
{0.25}&{0.5}\\
{0.5}&1
\end{array}} \right]
\end{equation}
Motion parameters of targets are shown in Table \ref{motionparameters}, initial state of each target obeys Gaussian distribution, and initial state covariance is set as ${\bf{Q}}$. At each time step, the measurement is a $128 \times 128$ image consisting of an array of cells with a scalar intensity, giving each cell side lengths of $\Delta x = \Delta y = 1$. The array index is treated as an ordered pair of integers $n = \left( {i,j} \right),1 \le i,j \le 128$. The point spread function in equation (\ref{measurementequation}) is
\begin{equation}
{h^n}\left( {\bf{x}} \right) = I\left( {\bf{x}} \right) \times \exp \left( { - \frac{{{{\left( {i\Delta x - x} \right)}^2} + {{\left( {j\Delta y - y} \right)}^2}}}{{\sigma _h^2}}} \right)
\end{equation}
where $I\left( {\bf{x}} \right)$ is the constant intensity of target ${\bf{x}}$, being $10$, $7$, $8$, $9$, and $10$ for the five targets respectively, and $\sigma _h^2 = 2$ is blurring factor. When ${h^n}\left( {\bf{x}} \right)$ is greater than preset threshold ${h^{{\rm{th}}}} = 1$, it is considered that cell $n$ is illuminated by target ${\bf{x}}$. The covariance of measurement noise ${w^n}$ in equation (\ref{measurementequation}) is $R = 1$.

To ensure a fair comparison, the multi-Bernoulli-TBD filter \cite{Vo2010} is implemented by sigma points transformation, and it shares the same parameters with the proposed target clustering based multi-Bernoulli filter for superpositional sensors. The survival probability is ${p_S} = 0.99$. At each time step, target with posterior existence probability greater than ${r^{{\rm{th}}}} = 0.99$ is extracted, and that with posterior existence probability less than ${r^{\rm{p}}} = {10^{ - 4}}$ is pruned. The number of sigma points is ${N_S} = 9$, and sigma points and weights are determined using \cite{Julier2004}. In the Gaussian implementation of the proposed TC-MB filter, the set of resolution cells illuminated by prior target is determined by its point spread function and preset threshold ${h^{{\rm{th}}}}$, which is used for target clustering.

Cardinality and Optimal Sub-Pattern Assignment (OSPA) \cite{Schuhmacher2008} distance between ground truth of multi-target state set and estimation of multi-target state set are utilized as performance evaluation criterions of filters, where the cut-off factor and the order used in OSPA are $c = 10$ and $p = 1$, respectively.

\begin{figure}
\centerline{\includegraphics[width=8cm]{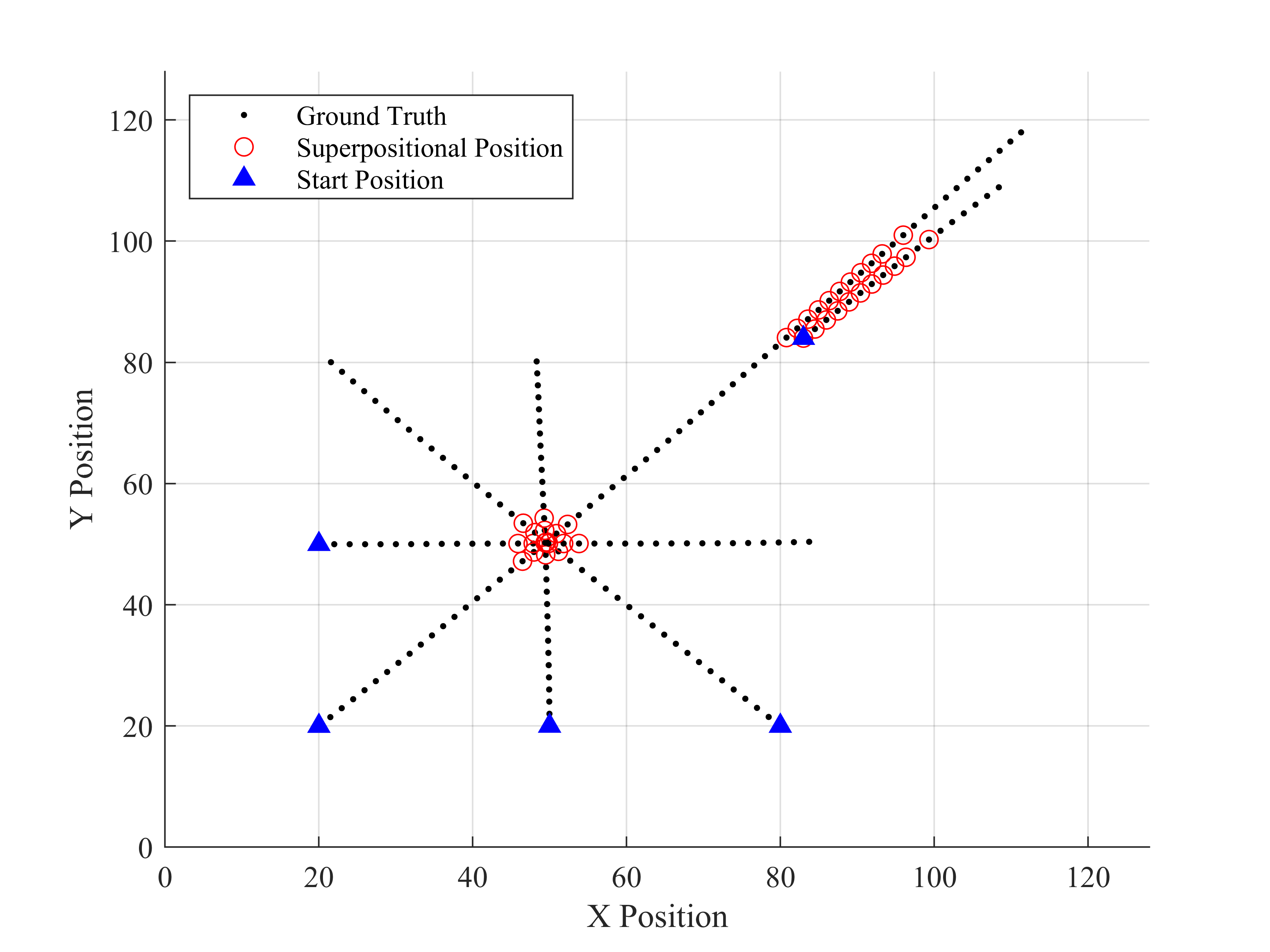}}
\caption{Simulation scenario of one trial\label{scenario}}
\end{figure}

Simulation scenario of one trial is depicted in Fig.\ref{scenario}. The resolution cell is called superpositional position (the red circle) if it is illuminated by multiple targets simultaneously. Near $k = 21$, target 1, 2, 3, and 4 almost reach the same cell (almost cell $\left( {50,50} \right)$) at the same time. In addition, the resolution cells illuminated by target 1 and 5 are overlapping for a long time.

\begin{figure}
\centerline{\includegraphics[width=8cm]{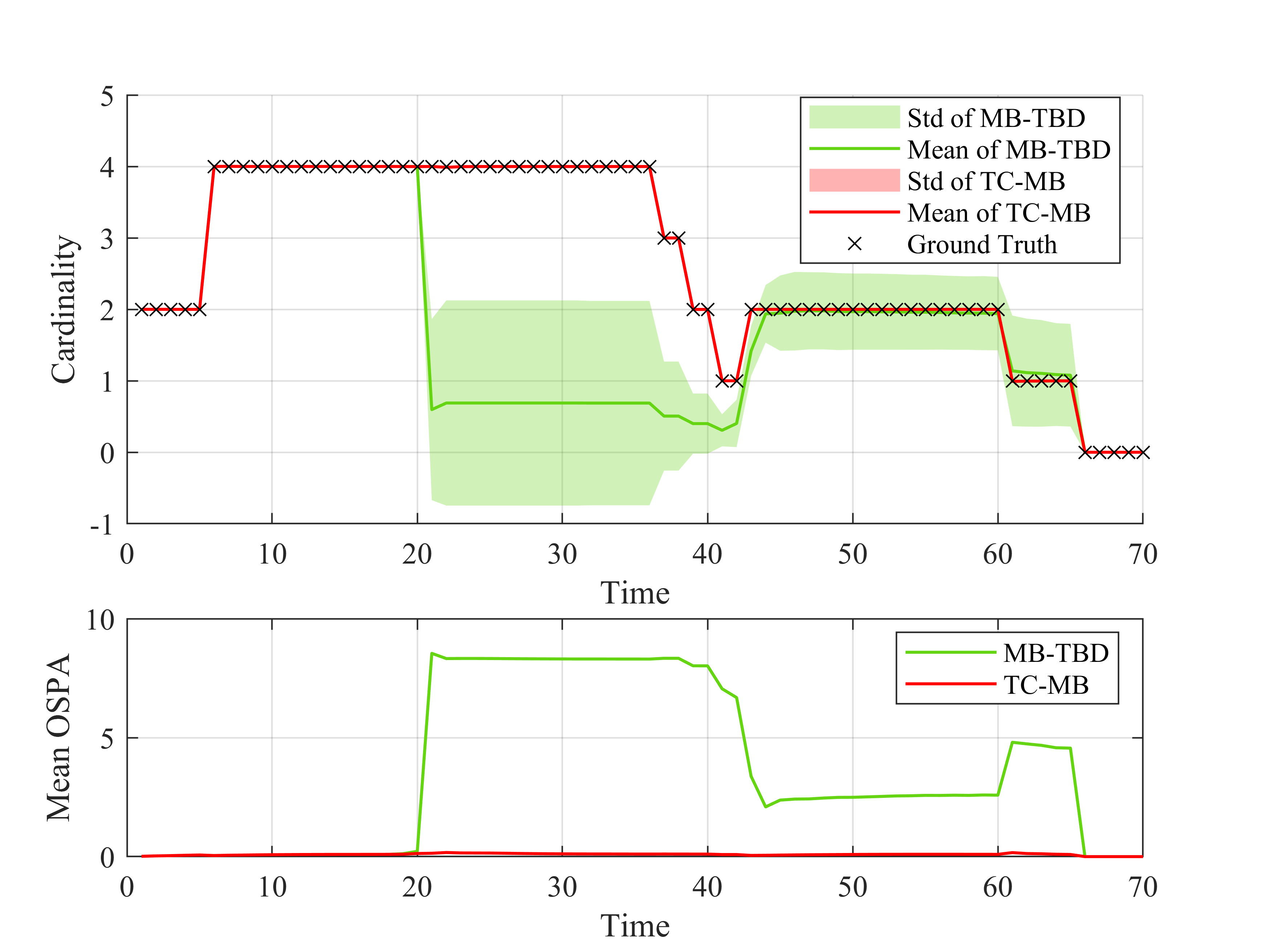}}
\caption{Cardinality and OSPA performances of MB-TBD filter and TC-MB filter\label{Cardinality_OSPA}}
\end{figure}

Fig.\ref{Cardinality_OSPA} depicts cardinality and OSPA performances of MB-TBD filter and TC-MB filter versus time over $1000$ Monte Carlo trials. Due to the interaction of four targets at $k = 21$, mean cardinality of MB-TBD filter drops dramatically and corresponding mean OSPA deteriorates to $8$ until target 2,3, and 4 die. In the second event, long term overlapping of two targets significantly deteriorates mean OSPA of MB-TBD filter and increases standard deviation (std) of its cardinality. The proposed TC-MB filter performs well throughout the simulation.

\begin{table}
\caption{Motion Parameters of Targets}
\label{motionparameters}
\begin{center}
\begin{tabular}{p{1cm}p{2.5cm}p{1.35cm}p{1.35cm}}
\hline
\hline
\rule[-1ex]{0pt}{3.5ex}Target&Mean of Initial State&Birth Time&Death Time\\
\hline
\rule[-1ex]{0pt}{3.5ex}1&$\left[ {20;1.5;20;1.5} \right]$&1&65\\
\hline
\rule[-1ex]{0pt}{3.5ex}2&$\left[ {80; - 1.5;20;1.5} \right]$&1&40\\
\hline
\rule[-1ex]{0pt}{3.5ex}3&$\left[ {20;2;50;0} \right]$&6&38\\
\hline
\rule[-1ex]{0pt}{3.5ex}4&$\left[ {50;0;20;2} \right]$&6&36\\
\hline
\rule[-1ex]{0pt}{3.5ex}5&$\left[ {83;1.5;84;1.5} \right]$&43&60\\
\hline
\hline
\end{tabular}
\end{center}
\end{table}

\section{Conclusion}
In this letter, target clustering based multi-Bernoulli filter for superpositional sensors is proposed. Single target posterior density is strictly derived, and multi-target posterior is approximated by the product of multiple Bernoulli densities. The Gaussian implementation of the proposed algorithm is also presented. The simulation results illustrate that the proposed algorithm can effectively track multiple targets when raw measurements are superpositional.

\bibliography{MBSS}
\end{document}